\documentclass[
reprint,
superscriptaddress,
 amsmath,amssymb,
prl,
floatfix,
]{revtex4-2}
\usepackage{dcolumn}
\usepackage{bm}
\usepackage{turnstile}
\usepackage{float}
\usepackage{graphicx}
\usepackage{amsmath,amssymb}
\usepackage{url}
\usepackage{epsfig}

\begin{document}

\title{
Tunable spin-flop transition in artificial ferrimagnets
}
\author{N.~O.~Antropov}
\affiliation{Institute of Metal Physics, 620180 Ekaterinburg, Russia}
\affiliation{Ural Federal University, 620002 Ekaterinburg, Russia}
\author{E.~A.~Kravtsov}
\affiliation{Institute of Metal Physics, 620180 Ekaterinburg, Russia}
\affiliation{Ural Federal University, 620002 Ekaterinburg, Russia}
\author{M.~V.~Makarova}
\affiliation{Institute of Metal Physics, 620180 Ekaterinburg, Russia}
\affiliation{Ural Federal University, 620002 Ekaterinburg, Russia}
\author{V.~V.~Proglyado}
\affiliation{Institute of Metal Physics, 620180 Ekaterinburg, Russia}
\author{T.~Keller}
\affiliation{Max-Planck-Institut f\"ur Festk\"orperforschung, Heisenbergstra\ss e 1, D-70569 Stuttgart, Germany}
\affiliation{Max Planck Society Outstation at the Heinz Maier-Leibnitz Zentrum (MLZ), D-85748 Garching, Germany}
\author{I.~A.~Subbotin}
\affiliation{National Research Center "Kurchatov Institute", 123182 Moscow, Russia}
\author{E.~M.~Pashaev}
\affiliation{National Research Center "Kurchatov Institute", 123182 Moscow, Russia}
\author{G.~V.~Prutskov}
\affiliation{National Research Center "Kurchatov Institute", 123182 Moscow, Russia}
\author{A.~L.~Vasiliev}
\affiliation{National Research Center "Kurchatov Institute", 123182 Moscow, Russia}
\author{Yu.~M.~Chesnokov}
\affiliation{National Research Center "Kurchatov Institute", 123182 Moscow, Russia}
\author{N.~G.~Bebenin}
\affiliation{Institute of Metal Physics, 620180 Ekaterinburg, Russia}
\author{V.~V.~Ustinov}
\affiliation{Institute of Metal Physics, 620180 Ekaterinburg, Russia}
\author{B.~Keimer}
\affiliation{Max-Planck-Institut f\"ur Festk\"orperforschung, Heisenbergstra\ss e 1, D-70569 Stuttgart, Germany}
\author{Yu.~N.~Khaydukov}
\affiliation{Max-Planck-Institut f\"ur Festk\"orperforschung, Heisenbergstra\ss e 1, D-70569 Stuttgart, Germany}
\affiliation{Max Planck Society Outstation at the Heinz Maier-Leibnitz Zentrum (MLZ), D-85748 Garching, Germany}
\affiliation{Skobeltsyn Institute of Nuclear Physics, Moscow State University, Moscow 119991, Russia}
\date{\today}

\begin{abstract}
Spin-flop transition (SFT) consists in a jump-like reversal of antiferromagnetic magnetic moments into a non-collinear state when the magnetic field increases above the critical value. Potentially the SFT can be utilized in many applications of a rapidly developing antiferromagnetic spintronics.  However, the difficulty of using them in conventional antiferromagnets lies in (a) too large switching magnetic fields (b) the need for presence of a magnetic anisotropy, and (c) requirement to apply magnetic field along the correspondent anisotropy axis. In this work we propose to use artificial ferrimagnets in which the spin-flop transition occurs without anisotropy and the transition field can be lowered by adjusting exchange coupling in the structure. This is proved by experiment on artificial Fe-Gd ferrimagnets where usage of Pd spacers allowed us to suppress the transition field by two orders of magnitude.

  \end{abstract}

\maketitle
 Antiferromagnetic (AF) spintronic is nowadays a rapidly developing area \cite{McDonald11,Jungwirth18,Duine18,Hirohata20,Shi20}. In addition to non-volatility of conventional ferromagnetic spintronics the AF devices can offer immunity to external magnetic disturbances, absence of cross-talks between small-area devices and much faster dynamics (THz vs MHz). The antiferromagnetic systems are featured by spin-flop transition (SFT) when there is the transition from antiferromagnetic ordering to noncollinear (NC) state at  magnetic field exceeding certain value $H_{SP}$. Creation of noncollinear magnetic state and possibility to switch between AF and NC states may have useful applications by utilizing anomalous Hall or Nernst effects \cite{Chen14,Nakatsuji15,Hoffman16,Hoffman18,Qin19,Yang20}. In addition, proximity of noncollinear magnetic texture to superconducting layer generates long-range triplet superconductivity which may also find diverse applications in superconducting spintronics \cite{Bergeret01,Volkov03,eschrig2011,KlosePRL12,LenkPRB17}. 
 
 The utilization of the spin-flop effect in AF systems is overly complicated due to at least two reasons. The first thing is the existence of SFT in AF requires uniaxial anisotropy and an external field applied along the corresponding axis. Secondly, typical transition fields $H_{SP}$ in bulk antiferromagnets are tens of Tesla \cite{Yokosuk16,Machado17,Becker17,Vibhakar19} thus they are too high for real applications. The need to have anisotropy inside the system can be circumvented by replacing antiferromagnets with ferrimagnets (FEMs). In the FEMs one does not require presence of anisotropy and the SFT takes place at $H_{SP}=\lambda|m_1-m_2|$ \cite{Clark1968}, where $m_{1,2}$ are the magnetic moment of first and second sublattices and $\lambda$ is the exchange parameter. In bulk systems the $H_{SP}$ are still too high for applications and can hardly be tuned.

 In contrast, artificial ferrimagnets based on  magnetic heterostructures give a possibility to tune the SFT field by varying parameters of ferromagnetic layers and by introducing non-magnetic spacers. Heterostructures based on 3d transition metals (TM) and heavy 4f rare-earth (RE) metals, like Fe/Gd, are model ferrimagnetic systems demonstrating a rich magnetic phase diagram with complex types of magnetic ordering \cite{IshimatsuPRB1999,HaskelPRL2001,Montoya1PRB17,Montoya2PRB17,Takanashi1992,Baczewski}. Coupling between 4f electrons of Gd and 3d electrons of Fe leads to the antiferromagnetic alignment of TM and RE magnetic moments which due to the difference in magnetic moments of Fe($\sim 2\mu_B$) and Gd ($\sim 7\mu_B$) leads to the emergence of a one-dimensional ferrimagnetic lattice.  The spin-flop transition was found in Gd/Fe systems at typical value $H_{SP}\sim$3kOe \cite{Kamiguchi89}, which is much smaller than that for bulk FEMs but still quite high for applications. Further tuning of $H_{SP}$ can be  gained by suppression of interlayer exchange coupling which can be performed by  spacing of Fe and Gd with a non-magnetic material like Cr \cite{Drovosekov15,Drovosekov18}, Pt \cite{Takanashi1993} or Si \cite{Merenkov2001}.

The SFT can be detected by integral magnetic techniques as a kink on a magnetic hysteresis loop at $H_{SP}$. In case of artificial FEMs magnetic signal from thin films is  heavily polluted by dia- or paramagnetic signal of thick substrates.This makes it difficult, if not impossible at all, to use integral magnetometric methods to study the SFTs. Neutron scattering, being a depth-selective magnetometric method is a widely used method for studying AFs and FEMs  \cite{Velthuis02,Pasyuk02,Nagy02}. Similar to X-ray and light, neutrons diffract at periodic lattice with period $D$ according to the well-known Bragg law $n\lambda = 2D\sin\theta$. Here $\lambda$ and $\theta$ are the neutron wavelength and incident angle, and $n$ is integer number corresponding to order of Bragg peak. Presence of spin one-half makes neutron scattering sensitive to the magnetic lattice. In case of antiferromagnetic lattice  magnetic peak is doubled comparing to the structural one, so that the magnetic Bragg peak appears on the positions of $n/2$ of the structural Bragg peaks. Applying spin analysis, that is detecting neutron spin-states before and after scattering, allows one to get additional information about magnetic configuration. The non-spin-flip (NSF) channels (++) and (- -) are sensitive to the sum and difference of nuclear potential and collinear to the neutron polarization part of magnetization. Here first and second sign codes neutron polarization along the external magnetic field $H$ before and after the scattering process. Presence of non-collinear magnetization causes spin-flip (SF) scattering (+-) and (-+). In Born approximation the amplitude of the SF scattering is proportional to the spatial profile of the noncollinear magnetization in reciprocal space. Thus the SF scattering is very sensitive channel to detect the SFTs.  

 In our prior work \cite{ANTROPOV} we studied superlattice [Fe(3.5nm)/Pd(1.2nm)/Gd(5nm)/Pd(1.2nm)]$_{12}$. In the neutron experiment we measured intensity of SF scattering at the position of the first Bragg peak $R^{SF}_1$  as a function of external magnetic field at a temperature of 10K. Above magnetic field of $H_{SP}$=1.5kOe we detected a 20-fold increase of SF scattering which is the direct evidence for the presence of SFT in our system. We note that the $H_{SP}$ field is much smaller than in spacer free Fe/Gd systems. Subsequent structural studies by transmission electron microscopy and synchrotron radiation \cite{Pashaev20} indicated presence of mutual diffusion at Gd/Pd interface. For thin ($\sim$1nm) Pd spacers this interdiffusion leads to almost complete dissolution of Pd in Gd. As a result the Curie temperature (and hence exchange energy) of the (nominal) Gd layer decreases from 294K for bulk Gd to $\lesssim$ 100K. Thus ability of Pd and Gd to form an alloy with controllable suppression of exchange energy paves the way for tuning of SFT by varying thickness of Pd spacer. To do this we prepared series of samples of nominal composition [Fe(3.5nm)/Pd(t)/Gd(5.0nm)/Pd(t)]$_{12}$ varying $t$ from 1.0 to 1.6 nm (details can be found in our prior works \cite{ANTROPOV, Pashaev20}). Further we will code samples as PdYY, where YY is thickness of Pd layer in Angstroms. 
 
\begin{figure}[htb]
	\centering
	\includegraphics[width=\columnwidth]{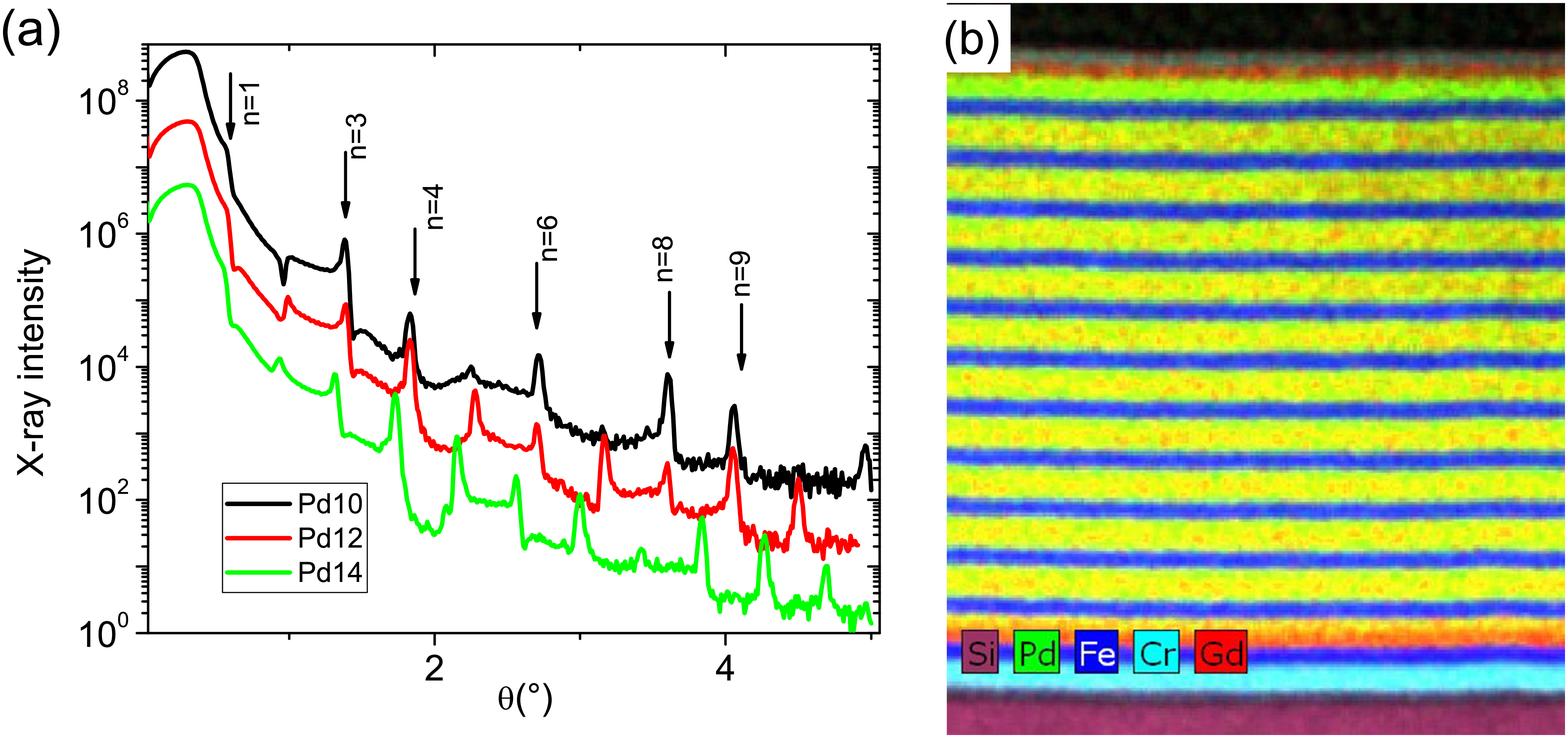}
	\caption{
 (a) X-ray low-angle diffraction (reflectivity) of samples under study. Vertical arrows show the position of several Bragg peaks for sample Pd10. (b) The energy dispersive X-ray (EDX) microanalysis of Pd12 sample.
	}
	\label{Qual}
\end{figure}

  Fig. \ref{Qual}a shows the X-ray low-angle diffraction patterns (reflectivities) measured at a wavelength of $\lambda$=1.54\AA~from the samples under study.  More than 10 orders of Bragg reflection are seen on the reflectivities, which indicates good repeatability of the Fe/Gd unit cell. Fig. \ref{Qual}b shows the energy dispersive X-ray (EDX) microanalysis of scanning transmission electron microscopy (STEM) of Pd12 sample. The EDX analysis shows well-defined Fe layers depicted by blue color and yellow layers of GdPd alloy instead of separate red Gd layers and green Pd spacers. For the sake of simplicity, we will keep naming Gd layer, remembering however that in reality the layer is a Gd$_x$Pd$_{1-x}$ alloy.

\begin{figure}[htb]
	\centering
	\includegraphics[width=1.0\columnwidth]{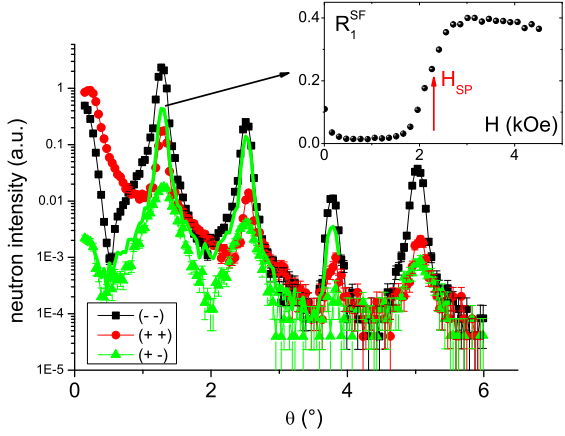}
	\caption{
		Polarized neutron reflectivities of sample Pd10 measured at $T=10$ K at magnetic field $H=1$ kOe (symbols) and SF curve at $T$=10 K, $H$=3kOe (solid line) Inset shows the field dependence of intensity of SF scattering at the first Bragg peak $R^{SF}_1(H)$. Vertical arrow denotes the magnetic field at which spin-flop transition takes place.
	}
	\label{Pd10}
\end{figure}

Polarized neutron reflectometry (PNR) experiment was conducted on the monochromatic ($\lambda$=4.3\AA) reflectometer NREX of the research reactor FRM-2 (Garching, Germany).  Fig.\ref{Pd10} shows the PNR data measured on sample Pd10 at $T$=10 K in magnetic field $H$=1kOe and additional SF curve at $T$=10 K in magnetic field $H$=3kOe (solid line). In the neutron experiment 4 Bragg peaks were confidently measured.  A large splitting of (++) and (-\,-) NSF Bragg peaks  indicates the presence of a collinear magnetic moment in the system. At the same time we observed a much weaker (1-2 orders below NSF signal) SF scattering at Bragg peaks. The origin of this small, though not negligible SF signal can be associated with noncollinear inhomogeneities at the Fe/Gd interfaces. The data at $H$=1kOe can be quantitatively described by a predominantly collinear AF state with magnetic moments of Gd $M_{Gd} \approx 5\mu_B$ and Fe $M_{Fe} \approx 2\mu_B$ aligned parallel and antiparallel to $H$.  By increasing the magnetic field above $H_{SP}$=2.3kOe (inset in Fig.\ref{Pd10}) we observed a 20-fold increase of SF scattering at the first Bragg peak $R^{SF}_1$. This SFT is similar to observed previously spin-flop in Pd12 sample though taking place at 1kOe higher magnetic field.

\begin{figure}[htb]
	\centering
	\includegraphics[width=1\columnwidth]{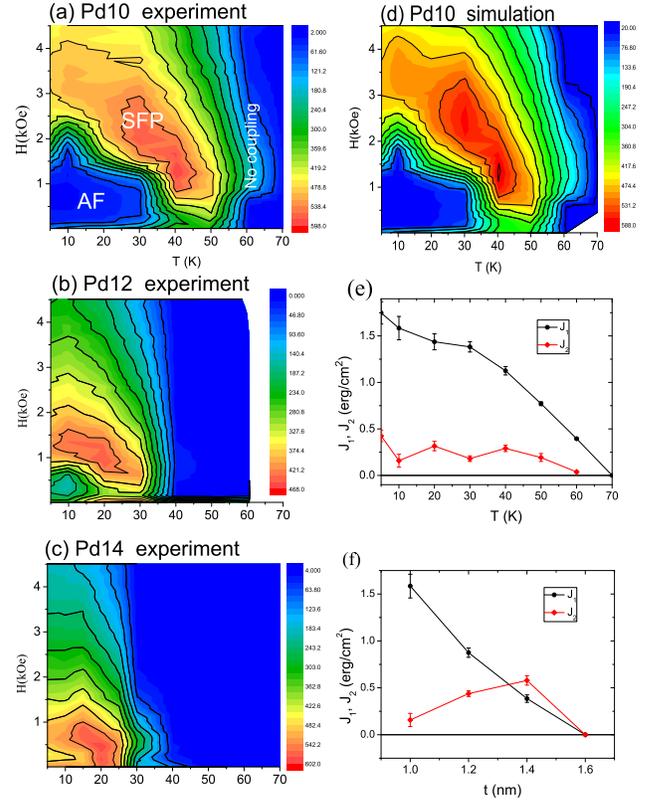}
	\caption{(a)-(c) Experimental ($H$,$T$) maps of $R^{SF}_1$ for samples with different Pd spacer. (d) Simulated map for Pd10 sample  (e) Fit-resulted $J_{1}$ and $J_{2}$ terms vs temperature for Pd10 sample. (f) Thickness dependence of bilinear and biquadratic energies $J_{1}$ and $J_{2}$ obtained for $T$=10K.
	}
	\label{Main}
\end{figure}

By measuring family of $R^{SF}_1$(H) scans at different temperatures we were able to construct the noncollinear magnetic phase diagram for the sample Pd10 in $H$-$T$ coordinates (Fig. \ref{Main}a). For this sample we observe a collinear AF state in the temperature range up to 30 K  in magnetic fields not exceeding 2 kOe. Above this field, the collinear AF state is replaced by a NC spin-flop state. Increasing the temperature to 60K leads to a gradual shift of the SFT field towards lower values. Finally, above 60K, the spin-flip signal disappears due to the absence of magnetic ordering in Gd layer. Fig.\ref{Main}b and Fig.\ref{Main}c shows similar phase diagrams for Pd12 and Pd14 samples.  One can see that the transition field $H_{SP}$ decreases with increase of $t$. For the samples with $t$=1.6nm (not shown) we did not observe any detectable SF signal evidencing absence of coupling of Fe and Gd layers.    

To describe magnetic state of our systems we applied extended Stoner-Wohlfarth model widely used for description of magnetic multilayers \cite{Solignac12,Hoffman16}.
Density of magnetic energy of one Fe/Gd unit cell can be written as

\begin{equation}\label{SW}
\begin{split}
E(\alpha_{Gd},\alpha_{Fe})=-H[m_{Gd}cos(\alpha_{Gd})+ m_{Fe}cos(\alpha_{Fe})] +\\
                            J_{1}cos(\alpha_{Gd}-\alpha_{Fe})+ J_{2}cos^{2}(\alpha_{Gd}-\alpha_{Fe}). \\
\end{split}
\end{equation}
In Eq.\ref{SW}  $m_{X}=M_{X}d_{X}$ is a product of magnetization and thickness (magnetic moment), $\alpha_{X}$ is the angle between magnetization and $H$ of a layer $X$ ($X$=Fe,Gd). The first term in (\ref{SW}) is Zeeman coupling which tends to align magnetic moments of the layers along the external field. The second term is bilinear antiferromagnetic exchange coupling of Fe and Gd layers with strength parameter $J_{1}$. The third term describes biquadratic coupling tending to align the magnetic moments non-collinearly. As seen from (\ref{SW}) in case $J_2$=0 the transition field can be estimated as $H_{SP}\approx J_{1}|m_{Gd}-m_{Fe}|/m_{Gd}\cdot m_{Fe}$.

For every magnetic field $H$ the  magnetic configuration of the system as a function of $J_{1,2}$ can be obtained by minimizing energy (\ref{SW}) varying angles $\alpha_{Gd}$ and $\alpha_{Fe}$. The magnetization amplitudes $M_{Gd, Fe}$ and thicknesses $d_{Gd, Fe}$ were taken from PNR and SQUID data and fixed during calculations. The angles $\alpha^{'}_{Gd}$ and $\alpha^{'}_{Fe}$ corresponding to the minimum of energy for a given set of $H$ and $J_{1,2}$ is used to construct a theoretical SF reflectivity at the first Bragg peak in Born approximation:

\begin{equation}
\begin{split}
\label{BA}
R^{SF}_{1,th}=c[m_{Gd,\bot}^2 +m_{Fe,\bot}^2+ \\
2m_{Gd,\bot}m_{Fe,\bot}\cos\frac{d_{Fe}}{d_{Fe}+d_{Gd}}] +R_{bg}, 
\end{split}
\end{equation}
where $m_{Gd(Fe),\bot}=m_{Gd(Fe)}\sin\alpha^{'}_{Gd(Fe)}$ is the non-collinear component of magnetic moment of Gd(Fe) layer, $c$ is scaling constant and $R_{bg}$ is background intensity. The latter two values were adjusted manually before the fit. We fitted then theoretical $R^{SF}_{1,th}$ to the experimental $H$-dependencies $R^{SF}_{1}$ by varying $J_1$ and $J_2$.  The procedure was repeated for every $T$ so that for every sample we obtained temperature dependencies of $J_{1,2}$.  Fig.\ref{Main}d shows results of such a fit for sample Pd10. It is rather noticeable that despite of the simplicity of the Stoner-Wohlfarth approach it allows to reproduce experimental features quite well. Fig.\ref{Main}e shows the fit-resulted $T$-dependence of the exchange energies $J_{1}$ and $J_{2}$ for Pd10 sample. It can be seen that the bilinear term has a predominant contribution, which gradually decreases with decreasing temperature. Thus our analysis showed that for a qualitative description of the SFT, a bilinear term is sufficient, but quantitatively the data are described better by including an additional biquadratic term.

The data for the other samples were fitted in a similar way. Fig.\ref{Main}f shows the dependency of coupling energies on thickness of Pd spacer. As follows from the figure, the bilinear energy decreases almost linearly from 1.5 erg/cm$^2$ at $t$=1nm to 0 at $t$=1.6nm. Biquadratic energy in turn increases with $t$. The obtained values are of the same orders as $J_{1}\sim$ 0.8 erg/cm$^2$ and $J_{2}\sim$ 0.2 erg/cm$^2$ obtained in Ref.\cite{Suciu02} for Gd/Pt/Co multilayers at $T$=10K.

The decrease in the bilinear component with the increase in $t$ can obviously be correlated with a decrease in the effective concentration of Gd in the GdPd layer. At the same time, structural studies carried out earlier \cite{Pashaev20} indicate an increase in structural inhomogeneities with increasing of $t$ . It seems prudent to correlate this growth with an increase in the biquadratic component.

In conclusion, using PNR we performed a systematic study of magnetic configuration of [Fe(3.5nm)/Pd(t)/Gd(5.0nm)/Pd(t)]$_{12}$ heterostructures with t=1.0-1.6nm. By measuring neutron spin-flip scattering we have detected presence of magnetically non-collinear state at temperatures $T \lesssim$ 50 K in magnetic fields of above $H>$500 Oe for the samples with 1nm$<t<$1.4nm. By using of an extended Stoner-Wohlfarth model we were able to describe the observed transition as a competition of Zeeman energy, bilinear interaction of order of 1 erg/cm$^2$ and biquadratic addition of order of 0.5 erg/cm$^2$. The coupling energies can be tuned by varying thickness of spacer between 1nm and 1.4nm leading to the shift of the transition field below kilo-Oersted range. Our study opens perspectives for a purposeful design of artificial FEMs with adjustable field of spin-flop transition. Thus, the FEMs systems with low Curie temperature components studied in this work can be used in superconducting spintronics for generation of triplet superconductivitiy. An additional advantage here is the good compatibility of gadolinium with superconducting niobium \cite{KhaydukovPRB18,Khaydukov19}. For the room temperature applications one can use well-studied synthetic AFs such as Fe/Cr \cite{Pasyuk02,Velthuis02,Nagy02}, Fe/V \cite{Hjoravrson97,Leiner02} or Co/Cu \cite{Schreyer93,Hecker05}   where subsequent adjustment can be carried out by tuning of the coupling energy and the imbalance of the magnetic moments of the sub-lattices. 

We would like to thank M.A. Milyaev for assistance in preparation of the samples, A.B. Drovosekov and D.I. Kholin for fruitful discussion of the results. This work is partially based on experiments performed at the NREX instrument operated by the Max Planck Society at the MLZ), Garching, Germany, and supported by the Deutsche Forschungsgemeinschaft (Project No. 107745057-TRR80). Research in Ekaterinburg was performed within the state assignment of Minobrnauki of Russia (theme "Spin" No. AAAA-A18-118020290104-2) and was partly supported by Russian Foundation for Basic Research (Project No. 19-02-00674). 

\bibliography{FePdGd_Refs}

\begin{thebibliography}{45}%
\makeatletter
\providecommand \@ifxundefined [1]{%
 \@ifx{#1\undefined}
}%
\providecommand \@ifnum [1]{%
 \ifnum #1\expandafter \@firstoftwo
 \else \expandafter \@secondoftwo
 \fi
}%
\providecommand \@ifx [1]{%
 \ifx #1\expandafter \@firstoftwo
 \else \expandafter \@secondoftwo
 \fi
}%
\providecommand \natexlab [1]{#1}%
\providecommand \enquote  [1]{``#1''}%
\providecommand \bibnamefont  [1]{#1}%
\providecommand \bibfnamefont [1]{#1}%
\providecommand \citenamefont [1]{#1}%
\providecommand \href@noop [0]{\@secondoftwo}%
\providecommand \href [0]{\begingroup \@sanitize@url \@href}%
\providecommand \@href[1]{\@@startlink{#1}\@@href}%
\providecommand \@@href[1]{\endgroup#1\@@endlink}%
\providecommand \@sanitize@url [0]{\catcode `\\12\catcode `\$12\catcode
  `\&12\catcode `\#12\catcode `\^12\catcode `\_12\catcode `\%12\relax}%
\providecommand \@@startlink[1]{}%
\providecommand \@@endlink[0]{}%
\providecommand \url  [0]{\begingroup\@sanitize@url \@url }%
\providecommand \@url [1]{\endgroup\@href {#1}{\urlprefix }}%
\providecommand \urlprefix  [0]{URL }%
\providecommand \Eprint [0]{\href }%
\providecommand \doibase [0]{https://doi.org/}%
\providecommand \selectlanguage [0]{\@gobble}%
\providecommand \bibinfo  [0]{\@secondoftwo}%
\providecommand \bibfield  [0]{\@secondoftwo}%
\providecommand \translation [1]{[#1]}%
\providecommand \BibitemOpen [0]{}%
\providecommand \bibitemStop [0]{}%
\providecommand \bibitemNoStop [0]{.\EOS\space}%
\providecommand \EOS [0]{\spacefactor3000\relax}%
\providecommand \BibitemShut  [1]{\csname bibitem#1\endcsname}%
\let\auto@bib@innerbib\@empty
\bibitem [{\citenamefont {MacDonald}\ and\ \citenamefont
  {Tsoi}(2011)}]{McDonald11}%
  \BibitemOpen
  \bibfield  {author} {\bibinfo {author} {\bibfnamefont {A.~H.}\ \bibnamefont
  {MacDonald}}\ and\ \bibinfo {author} {\bibfnamefont {M.}~\bibnamefont
  {Tsoi}},\ }\bibfield  {title} {\bibinfo {title} {Antiferromagnetic metal
  spintronics},\ }\href@noop {} {\bibfield  {journal} {\bibinfo  {journal}
  {Philos. Trans. R. Soc. A}\ }\textbf {\bibinfo {volume} {369}},\ \bibinfo
  {pages} {3098} (\bibinfo {year} {2011})}\BibitemShut {NoStop}%
\bibitem [{\citenamefont {Jungwirth}\ \emph {et~al.}(2018)\citenamefont
  {Jungwirth}, \citenamefont {Sinova}, \citenamefont {Manchon}, \citenamefont
  {Marti}, \citenamefont {Wunderlich},\ and\ \citenamefont
  {Felser}}]{Jungwirth18}%
  \BibitemOpen
  \bibfield  {author} {\bibinfo {author} {\bibfnamefont {T.}~\bibnamefont
  {Jungwirth}}, \bibinfo {author} {\bibfnamefont {J.}~\bibnamefont {Sinova}},
  \bibinfo {author} {\bibfnamefont {A.}~\bibnamefont {Manchon}}, \bibinfo
  {author} {\bibfnamefont {X.}~\bibnamefont {Marti}}, \bibinfo {author}
  {\bibfnamefont {J.}~\bibnamefont {Wunderlich}},\ and\ \bibinfo {author}
  {\bibfnamefont {C.}~\bibnamefont {Felser}},\ }\bibfield  {title} {\bibinfo
  {title} {The multiple directions of antiferromagnetic spintronics},\
  }\href@noop {} {\bibfield  {journal} {\bibinfo  {journal} {Nat. Phys.}\
  }\textbf {\bibinfo {volume} {14}},\ \bibinfo {pages} {200} (\bibinfo {year}
  {2018})}\BibitemShut {NoStop}%
\bibitem [{\citenamefont {Duine}\ \emph {et~al.}(2018)\citenamefont {Duine},
  \citenamefont {Lee}, \citenamefont {Parkin},\ and\ \citenamefont
  {Stiles}}]{Duine18}%
  \BibitemOpen
  \bibfield  {author} {\bibinfo {author} {\bibfnamefont {R.~A.}\ \bibnamefont
  {Duine}}, \bibinfo {author} {\bibfnamefont {K.-J.}\ \bibnamefont {Lee}},
  \bibinfo {author} {\bibfnamefont {S.~S.~P.}\ \bibnamefont {Parkin}},\ and\
  \bibinfo {author} {\bibfnamefont {M.~D.}\ \bibnamefont {Stiles}},\ }\bibfield
   {title} {\bibinfo {title} {Synthetic antiferromagnetic spintronics},\
  }\href@noop {} {\bibfield  {journal} {\bibinfo  {journal} {Nat. Phys.}\
  }\textbf {\bibinfo {volume} {14}},\ \bibinfo {pages} {217} (\bibinfo {year}
  {2018})}\BibitemShut {NoStop}%
\bibitem [{\citenamefont {Hirohata}\ \emph {et~al.}(2020)\citenamefont
  {Hirohata}, \citenamefont {Yamada}, \citenamefont {Nakatani}, \citenamefont
  {Prejbeanu}, \citenamefont {Di{\'{e}}ny}, \citenamefont {Pirro},\ and\
  \citenamefont {Hillebrands}}]{Hirohata20}%
  \BibitemOpen
  \bibfield  {author} {\bibinfo {author} {\bibfnamefont {A.}~\bibnamefont
  {Hirohata}}, \bibinfo {author} {\bibfnamefont {K.}~\bibnamefont {Yamada}},
  \bibinfo {author} {\bibfnamefont {Y.}~\bibnamefont {Nakatani}}, \bibinfo
  {author} {\bibfnamefont {I.-L.}\ \bibnamefont {Prejbeanu}}, \bibinfo {author}
  {\bibfnamefont {B.}~\bibnamefont {Di{\'{e}}ny}}, \bibinfo {author}
  {\bibfnamefont {P.}~\bibnamefont {Pirro}},\ and\ \bibinfo {author}
  {\bibfnamefont {B.}~\bibnamefont {Hillebrands}},\ }\bibfield  {title}
  {\bibinfo {title} {Review on spintronics: Principles and device
  applications},\ }\href@noop {} {\bibfield  {journal} {\bibinfo  {journal} {J.
  Magn. Magn. Mater.}\ }\textbf {\bibinfo {volume} {509}},\ \bibinfo {pages}
  {166711} (\bibinfo {year} {2020})}\BibitemShut {NoStop}%
\bibitem [{\citenamefont {Shi}\ \emph {et~al.}(2020)\citenamefont {Shi},
  \citenamefont {Lopez-Dominguez}, \citenamefont {Garesci}, \citenamefont
  {Wang}, \citenamefont {Almasi}, \citenamefont {Grayson}, \citenamefont
  {Finocchio},\ and\ \citenamefont {Amiri}}]{Shi20}%
  \BibitemOpen
  \bibfield  {author} {\bibinfo {author} {\bibfnamefont {J.}~\bibnamefont
  {Shi}}, \bibinfo {author} {\bibfnamefont {V.}~\bibnamefont
  {Lopez-Dominguez}}, \bibinfo {author} {\bibfnamefont {F.}~\bibnamefont
  {Garesci}}, \bibinfo {author} {\bibfnamefont {C.}~\bibnamefont {Wang}},
  \bibinfo {author} {\bibfnamefont {H.}~\bibnamefont {Almasi}}, \bibinfo
  {author} {\bibfnamefont {M.}~\bibnamefont {Grayson}}, \bibinfo {author}
  {\bibfnamefont {G.}~\bibnamefont {Finocchio}},\ and\ \bibinfo {author}
  {\bibfnamefont {P.~K.}\ \bibnamefont {Amiri}},\ }\bibfield  {title} {\bibinfo
  {title} {Electrical manipulation of the magnetic order in antiferromagnetic
  {PtMn} pillars},\ }\href@noop {} {\bibfield  {journal} {\bibinfo  {journal}
  {Nat. Electron.}\ }\textbf {\bibinfo {volume} {3}},\ \bibinfo {pages} {92}
  (\bibinfo {year} {2020})}\BibitemShut {NoStop}%
\bibitem [{\citenamefont {Chen}\ \emph {et~al.}(2014)\citenamefont {Chen},
  \citenamefont {Niu},\ and\ \citenamefont {MacDonald}}]{Chen14}%
  \BibitemOpen
  \bibfield  {author} {\bibinfo {author} {\bibfnamefont {H.}~\bibnamefont
  {Chen}}, \bibinfo {author} {\bibfnamefont {Q.}~\bibnamefont {Niu}},\ and\
  \bibinfo {author} {\bibfnamefont {A.~H.}\ \bibnamefont {MacDonald}},\
  }\bibfield  {title} {\bibinfo {title} {Anomalous hall effect arising from
  noncollinear antiferromagnetism},\ }\href@noop {} {\bibfield  {journal}
  {\bibinfo  {journal} {Phys. Rev. Lett.}\ }\textbf {\bibinfo {volume} {112}},\
  \bibinfo {pages} {017205} (\bibinfo {year} {2014})}\BibitemShut {NoStop}%
\bibitem [{\citenamefont {Nakatsuji}\ \emph {et~al.}(2015)\citenamefont
  {Nakatsuji}, \citenamefont {Kiyohara},\ and\ \citenamefont
  {Higo}}]{Nakatsuji15}%
  \BibitemOpen
  \bibfield  {author} {\bibinfo {author} {\bibfnamefont {S.}~\bibnamefont
  {Nakatsuji}}, \bibinfo {author} {\bibfnamefont {N.}~\bibnamefont
  {Kiyohara}},\ and\ \bibinfo {author} {\bibfnamefont {T.}~\bibnamefont
  {Higo}},\ }\bibfield  {title} {\bibinfo {title} {Large anomalous hall effect
  in a non-collinear antiferromagnet at room temperature},\ }\href@noop {}
  {\bibfield  {journal} {\bibinfo  {journal} {Nature}\ }\textbf {\bibinfo
  {volume} {527}},\ \bibinfo {pages} {212} (\bibinfo {year}
  {2015})}\BibitemShut {NoStop}%
\bibitem [{\citenamefont {Hoffman}\ \emph {et~al.}(2016)\citenamefont
  {Hoffman}, \citenamefont {Kirby}, \citenamefont {Kwon}, \citenamefont
  {Fabbris}, \citenamefont {Meyers}, \citenamefont {Freeland}, \citenamefont
  {Martin}, \citenamefont {Heinonen}, \citenamefont {Steadman}, \citenamefont
  {Zhou}, \citenamefont {Schlep\"utz}, \citenamefont {Dean}, \citenamefont
  {te~Velthuis}, \citenamefont {Zuo},\ and\ \citenamefont
  {Bhattacharya}}]{Hoffman16}%
  \BibitemOpen
  \bibfield  {author} {\bibinfo {author} {\bibfnamefont {J.~D.}\ \bibnamefont
  {Hoffman}}, \bibinfo {author} {\bibfnamefont {B.~J.}\ \bibnamefont {Kirby}},
  \bibinfo {author} {\bibfnamefont {J.}~\bibnamefont {Kwon}}, \bibinfo {author}
  {\bibfnamefont {G.}~\bibnamefont {Fabbris}}, \bibinfo {author} {\bibfnamefont
  {D.}~\bibnamefont {Meyers}}, \bibinfo {author} {\bibfnamefont {J.~W.}\
  \bibnamefont {Freeland}}, \bibinfo {author} {\bibfnamefont {I.}~\bibnamefont
  {Martin}}, \bibinfo {author} {\bibfnamefont {O.~G.}\ \bibnamefont
  {Heinonen}}, \bibinfo {author} {\bibfnamefont {P.}~\bibnamefont {Steadman}},
  \bibinfo {author} {\bibfnamefont {H.}~\bibnamefont {Zhou}}, \bibinfo {author}
  {\bibfnamefont {C.~M.}\ \bibnamefont {Schlep\"utz}}, \bibinfo {author}
  {\bibfnamefont {M.~P.~M.}\ \bibnamefont {Dean}}, \bibinfo {author}
  {\bibfnamefont {S.~G.~E.}\ \bibnamefont {te~Velthuis}}, \bibinfo {author}
  {\bibfnamefont {J.-M.}\ \bibnamefont {Zuo}},\ and\ \bibinfo {author}
  {\bibfnamefont {A.}~\bibnamefont {Bhattacharya}},\ }\bibfield  {title}
  {\bibinfo {title} {Oscillatory noncollinear magnetism induced by interfacial
  charge transfer in superlattices composed of metallic oxides},\ }\href@noop
  {} {\bibfield  {journal} {\bibinfo  {journal} {Phys. Rev. X}\ }\textbf
  {\bibinfo {volume} {6}},\ \bibinfo {pages} {041038} (\bibinfo {year}
  {2016})}\BibitemShut {NoStop}%
\bibitem [{\citenamefont {Hoffman}\ \emph {et~al.}(2018)\citenamefont
  {Hoffman}, \citenamefont {Wu}, \citenamefont {Kirby},\ and\ \citenamefont
  {Bhattacharya}}]{Hoffman18}%
  \BibitemOpen
  \bibfield  {author} {\bibinfo {author} {\bibfnamefont {J.~D.}\ \bibnamefont
  {Hoffman}}, \bibinfo {author} {\bibfnamefont {S.~M.}\ \bibnamefont {Wu}},
  \bibinfo {author} {\bibfnamefont {B.~J.}\ \bibnamefont {Kirby}},\ and\
  \bibinfo {author} {\bibfnamefont {A.}~\bibnamefont {Bhattacharya}},\
  }\bibfield  {title} {\bibinfo {title} {Tunable noncollinear antiferromagnetic
  resistive memory through oxide superlattice design},\ }\href
  {https://doi.org/10.1103/PhysRevApplied.9.044041} {\bibfield  {journal}
  {\bibinfo  {journal} {Phys. Rev. Applied}\ }\textbf {\bibinfo {volume} {9}},\
  \bibinfo {pages} {044041} (\bibinfo {year} {2018})}\BibitemShut {NoStop}%
\bibitem [{\citenamefont {Qin}\ \emph {et~al.}(2019)\citenamefont {Qin},
  \citenamefont {Yan}, \citenamefont {Wang}, \citenamefont {Feng},
  \citenamefont {Guo}, \citenamefont {Zhou}, \citenamefont {Wu}, \citenamefont
  {Zhang}, \citenamefont {Leng}, \citenamefont {Chen},\ and\ \citenamefont
  {Liu}}]{Qin19}%
  \BibitemOpen
  \bibfield  {author} {\bibinfo {author} {\bibfnamefont {P.-X.}\ \bibnamefont
  {Qin}}, \bibinfo {author} {\bibfnamefont {H.}~\bibnamefont {Yan}}, \bibinfo
  {author} {\bibfnamefont {X.-N.}\ \bibnamefont {Wang}}, \bibinfo {author}
  {\bibfnamefont {Z.-X.}\ \bibnamefont {Feng}}, \bibinfo {author}
  {\bibfnamefont {H.-X.}\ \bibnamefont {Guo}}, \bibinfo {author} {\bibfnamefont
  {X.-R.}\ \bibnamefont {Zhou}}, \bibinfo {author} {\bibfnamefont {H.-J.}\
  \bibnamefont {Wu}}, \bibinfo {author} {\bibfnamefont {X.}~\bibnamefont
  {Zhang}}, \bibinfo {author} {\bibfnamefont {Z.-G.-G.}\ \bibnamefont {Leng}},
  \bibinfo {author} {\bibfnamefont {H.-Y.}\ \bibnamefont {Chen}},\ and\
  \bibinfo {author} {\bibfnamefont {Z.-Q.}\ \bibnamefont {Liu}},\ }\bibfield
  {title} {\bibinfo {title} {Noncollinear spintronics and electric-field
  control: a review},\ }\href@noop {} {\bibfield  {journal} {\bibinfo
  {journal} {Rare Metals}\ }\textbf {\bibinfo {volume} {39}},\ \bibinfo {pages}
  {95} (\bibinfo {year} {2019})}\BibitemShut {NoStop}%
\bibitem [{\citenamefont {Yang}(2020)}]{Yang20}%
  \BibitemOpen
  \bibfield  {author} {\bibinfo {author} {\bibfnamefont {S.-H.}\ \bibnamefont
  {Yang}},\ }\bibfield  {title} {\bibinfo {title} {Spintronics on chiral
  objects},\ }\href@noop {} {\bibfield  {journal} {\bibinfo  {journal} {Appl.
  Phys. Lett.}\ }\textbf {\bibinfo {volume} {116}},\ \bibinfo {pages} {120502}
  (\bibinfo {year} {2020})}\BibitemShut {NoStop}%
\bibitem [{\citenamefont {Bergeret}\ \emph {et~al.}(2001)\citenamefont
  {Bergeret}, \citenamefont {Volkov},\ and\ \citenamefont
  {Efetov}}]{Bergeret01}%
  \BibitemOpen
  \bibfield  {author} {\bibinfo {author} {\bibfnamefont {F.~S.}\ \bibnamefont
  {Bergeret}}, \bibinfo {author} {\bibfnamefont {A.~F.}\ \bibnamefont
  {Volkov}},\ and\ \bibinfo {author} {\bibfnamefont {K.~B.}\ \bibnamefont
  {Efetov}},\ }\bibfield  {title} {\bibinfo {title} {Long-range proximity
  effects in superconductor-ferromagnet structures},\ }\href@noop {} {\bibfield
   {journal} {\bibinfo  {journal} {Phys. Rev. Lett.}\ }\textbf {\bibinfo
  {volume} {86}},\ \bibinfo {pages} {4096} (\bibinfo {year}
  {2001})}\BibitemShut {NoStop}%
\bibitem [{\citenamefont {Volkov}\ \emph {et~al.}(2003)\citenamefont {Volkov},
  \citenamefont {Bergeret},\ and\ \citenamefont {Efetov}}]{Volkov03}%
  \BibitemOpen
  \bibfield  {author} {\bibinfo {author} {\bibfnamefont {A.~F.}\ \bibnamefont
  {Volkov}}, \bibinfo {author} {\bibfnamefont {F.~S.}\ \bibnamefont
  {Bergeret}},\ and\ \bibinfo {author} {\bibfnamefont {K.~B.}\ \bibnamefont
  {Efetov}},\ }\bibfield  {title} {\bibinfo {title} {Odd triplet
  superconductivity in superconductor-ferromagnet multilayered structures},\
  }\href@noop {} {\bibfield  {journal} {\bibinfo  {journal} {Phys. Rev. Lett.}\
  }\textbf {\bibinfo {volume} {90}},\ \bibinfo {pages} {117006} (\bibinfo
  {year} {2003})}\BibitemShut {NoStop}%
\bibitem [{\citenamefont {Eschrig}(2011)}]{eschrig2011}%
  \BibitemOpen
  \bibfield  {author} {\bibinfo {author} {\bibfnamefont {M.}~\bibnamefont
  {Eschrig}},\ }\bibfield  {title} {\bibinfo {title} {Spin-polarized
  supercurrents for spintronics},\ }\href@noop {} {\bibfield  {journal}
  {\bibinfo  {journal} {Phys. Today}\ }\textbf {\bibinfo {volume} {64}},\
  \bibinfo {pages} {43} (\bibinfo {year} {2011})}\BibitemShut {NoStop}%
\bibitem [{\citenamefont {Klose}\ \emph {et~al.}(2012)\citenamefont {Klose},
  \citenamefont {Khaire}, \citenamefont {Wang}, \citenamefont {Pratt},
  \citenamefont {Birge}, \citenamefont {McMorran}, \citenamefont {Ginley},
  \citenamefont {Borchers}, \citenamefont {Kirby}, \citenamefont {Maranville},\
  and\ \citenamefont {Unguris}}]{KlosePRL12}%
  \BibitemOpen
  \bibfield  {author} {\bibinfo {author} {\bibfnamefont {C.}~\bibnamefont
  {Klose}}, \bibinfo {author} {\bibfnamefont {T.~S.}\ \bibnamefont {Khaire}},
  \bibinfo {author} {\bibfnamefont {Y.}~\bibnamefont {Wang}}, \bibinfo {author}
  {\bibfnamefont {W.~P.}\ \bibnamefont {Pratt}}, \bibinfo {author}
  {\bibfnamefont {N.~O.}\ \bibnamefont {Birge}}, \bibinfo {author}
  {\bibfnamefont {B.~J.}\ \bibnamefont {McMorran}}, \bibinfo {author}
  {\bibfnamefont {T.~P.}\ \bibnamefont {Ginley}}, \bibinfo {author}
  {\bibfnamefont {J.~A.}\ \bibnamefont {Borchers}}, \bibinfo {author}
  {\bibfnamefont {B.~J.}\ \bibnamefont {Kirby}}, \bibinfo {author}
  {\bibfnamefont {B.~B.}\ \bibnamefont {Maranville}},\ and\ \bibinfo {author}
  {\bibfnamefont {J.}~\bibnamefont {Unguris}},\ }\bibfield  {title} {\bibinfo
  {title} {Optimization of spin-triplet supercurrent in ferromagnetic josephson
  junctions},\ }\href {https://link.aps.org/doi/10.1103/PhysRevLett.108.127002}
  {\bibfield  {journal} {\bibinfo  {journal} {Phys. Rev. Lett.}\ }\textbf
  {\bibinfo {volume} {108}},\ \bibinfo {pages} {127002} (\bibinfo {year}
  {2012})}\BibitemShut {NoStop}%
\bibitem [{\citenamefont {Lenk}\ \emph {et~al.}(2017)\citenamefont {Lenk},
  \citenamefont {Morari}, \citenamefont {Zdravkov}, \citenamefont {Ullrich},
  \citenamefont {Khaydukov}, \citenamefont {Obermeier}, \citenamefont
  {M\"uller}, \citenamefont {Sidorenko}, \citenamefont {von Nidda},
  \citenamefont {Horn}, \citenamefont {Tagirov},\ and\ \citenamefont
  {Tidecks}}]{LenkPRB17}%
  \BibitemOpen
  \bibfield  {author} {\bibinfo {author} {\bibfnamefont {D.}~\bibnamefont
  {Lenk}}, \bibinfo {author} {\bibfnamefont {R.}~\bibnamefont {Morari}},
  \bibinfo {author} {\bibfnamefont {V.~I.}\ \bibnamefont {Zdravkov}}, \bibinfo
  {author} {\bibfnamefont {A.}~\bibnamefont {Ullrich}}, \bibinfo {author}
  {\bibfnamefont {Y.}~\bibnamefont {Khaydukov}}, \bibinfo {author}
  {\bibfnamefont {G.}~\bibnamefont {Obermeier}}, \bibinfo {author}
  {\bibfnamefont {C.}~\bibnamefont {M\"uller}}, \bibinfo {author}
  {\bibfnamefont {A.~S.}\ \bibnamefont {Sidorenko}}, \bibinfo {author}
  {\bibfnamefont {H.-A.~K.}\ \bibnamefont {von Nidda}}, \bibinfo {author}
  {\bibfnamefont {S.}~\bibnamefont {Horn}}, \bibinfo {author} {\bibfnamefont
  {L.~R.}\ \bibnamefont {Tagirov}},\ and\ \bibinfo {author} {\bibfnamefont
  {R.}~\bibnamefont {Tidecks}},\ }\bibfield  {title} {\bibinfo {title}
  {Full-switching fsf-type superconducting spin-triplet magnetic random access
  memory element},\ }\href
  {https://link.aps.org/doi/10.1103/PhysRevB.96.184521} {\bibfield  {journal}
  {\bibinfo  {journal} {Phys. Rev. B}\ }\textbf {\bibinfo {volume} {96}},\
  \bibinfo {pages} {184521} (\bibinfo {year} {2017})}\BibitemShut {NoStop}%
\bibitem [{\citenamefont {Yokosuk}\ \emph {et~al.}(2016)\citenamefont
  {Yokosuk}, \citenamefont {al~Wahish}, \citenamefont {Artyukhin},
  \citenamefont {O'Neal}, \citenamefont {Mazumdar}, \citenamefont {Chen},
  \citenamefont {Yang}, \citenamefont {Oh}, \citenamefont {McGill},
  \citenamefont {Haule}, \citenamefont {Cheong}, \citenamefont {Vanderbilt},\
  and\ \citenamefont {Musfeldt}}]{Yokosuk16}%
  \BibitemOpen
  \bibfield  {author} {\bibinfo {author} {\bibfnamefont {M.~O.}\ \bibnamefont
  {Yokosuk}}, \bibinfo {author} {\bibfnamefont {A.}~\bibnamefont {al~Wahish}},
  \bibinfo {author} {\bibfnamefont {S.}~\bibnamefont {Artyukhin}}, \bibinfo
  {author} {\bibfnamefont {K.~R.}\ \bibnamefont {O'Neal}}, \bibinfo {author}
  {\bibfnamefont {D.}~\bibnamefont {Mazumdar}}, \bibinfo {author}
  {\bibfnamefont {P.}~\bibnamefont {Chen}}, \bibinfo {author} {\bibfnamefont
  {J.}~\bibnamefont {Yang}}, \bibinfo {author} {\bibfnamefont {Y.~S.}\
  \bibnamefont {Oh}}, \bibinfo {author} {\bibfnamefont {S.~A.}\ \bibnamefont
  {McGill}}, \bibinfo {author} {\bibfnamefont {K.}~\bibnamefont {Haule}},
  \bibinfo {author} {\bibfnamefont {S.-W.}\ \bibnamefont {Cheong}}, \bibinfo
  {author} {\bibfnamefont {D.}~\bibnamefont {Vanderbilt}},\ and\ \bibinfo
  {author} {\bibfnamefont {J.~L.}\ \bibnamefont {Musfeldt}},\ }\bibfield
  {title} {\bibinfo {title} {Magnetoelectric coupling through the spin flop
  transition in ${\mathrm{ni}}_{3}{\mathrm{teo}}_{6}$},\ }\href@noop {}
  {\bibfield  {journal} {\bibinfo  {journal} {Phys. Rev. Lett.}\ }\textbf
  {\bibinfo {volume} {117}},\ \bibinfo {pages} {147402} (\bibinfo {year}
  {2016})}\BibitemShut {NoStop}%
\bibitem [{\citenamefont {Machado}\ \emph {et~al.}(2017)\citenamefont
  {Machado}, \citenamefont {Ribeiro}, \citenamefont {Holanda}, \citenamefont
  {Rodr\'{\i}guez-Su\'arez}, \citenamefont {Azevedo},\ and\ \citenamefont
  {Rezende}}]{Machado17}%
  \BibitemOpen
  \bibfield  {author} {\bibinfo {author} {\bibfnamefont {F.~L.~A.}\
  \bibnamefont {Machado}}, \bibinfo {author} {\bibfnamefont {P.~R.~T.}\
  \bibnamefont {Ribeiro}}, \bibinfo {author} {\bibfnamefont {J.}~\bibnamefont
  {Holanda}}, \bibinfo {author} {\bibfnamefont {R.~L.}\ \bibnamefont
  {Rodr\'{\i}guez-Su\'arez}}, \bibinfo {author} {\bibfnamefont
  {A.}~\bibnamefont {Azevedo}},\ and\ \bibinfo {author} {\bibfnamefont {S.~M.}\
  \bibnamefont {Rezende}},\ }\bibfield  {title} {\bibinfo {title} {Spin-flop
  transition in the easy-plane antiferromagnet nickel oxide},\ }\href@noop {}
  {\bibfield  {journal} {\bibinfo  {journal} {Phys. Rev. B}\ }\textbf {\bibinfo
  {volume} {95}},\ \bibinfo {pages} {104418} (\bibinfo {year}
  {2017})}\BibitemShut {NoStop}%
\bibitem [{\citenamefont {Becker}\ \emph {et~al.}(2017)\citenamefont {Becker},
  \citenamefont {Tsukamoto}, \citenamefont {Kirilyuk}, \citenamefont {Maan},
  \citenamefont {Rasing}, \citenamefont {Christianen},\ and\ \citenamefont
  {Kimel}}]{Becker17}%
  \BibitemOpen
  \bibfield  {author} {\bibinfo {author} {\bibfnamefont {J.}~\bibnamefont
  {Becker}}, \bibinfo {author} {\bibfnamefont {A.}~\bibnamefont {Tsukamoto}},
  \bibinfo {author} {\bibfnamefont {A.}~\bibnamefont {Kirilyuk}}, \bibinfo
  {author} {\bibfnamefont {J.~C.}\ \bibnamefont {Maan}}, \bibinfo {author}
  {\bibfnamefont {T.}~\bibnamefont {Rasing}}, \bibinfo {author} {\bibfnamefont
  {P.~C.~M.}\ \bibnamefont {Christianen}},\ and\ \bibinfo {author}
  {\bibfnamefont {A.~V.}\ \bibnamefont {Kimel}},\ }\bibfield  {title} {\bibinfo
  {title} {Ultrafast magnetism of a ferrimagnet across the spin-flop transition
  in high magnetic fields},\ }\href@noop {} {\bibfield  {journal} {\bibinfo
  {journal} {Phys. Rev. Lett.}\ }\textbf {\bibinfo {volume} {118}},\ \bibinfo
  {pages} {117203} (\bibinfo {year} {2017})}\BibitemShut {NoStop}%
\bibitem [{\citenamefont {Vibhakar}\ \emph {et~al.}(2019)\citenamefont
  {Vibhakar}, \citenamefont {Khalyavin}, \citenamefont {Manuel}, \citenamefont
  {Zhang}, \citenamefont {Yamaura}, \citenamefont {Radaelli}, \citenamefont
  {Belik},\ and\ \citenamefont {Johnson}}]{Vibhakar19}%
  \BibitemOpen
  \bibfield  {author} {\bibinfo {author} {\bibfnamefont {A.~M.}\ \bibnamefont
  {Vibhakar}}, \bibinfo {author} {\bibfnamefont {D.~D.}\ \bibnamefont
  {Khalyavin}}, \bibinfo {author} {\bibfnamefont {P.}~\bibnamefont {Manuel}},
  \bibinfo {author} {\bibfnamefont {L.}~\bibnamefont {Zhang}}, \bibinfo
  {author} {\bibfnamefont {K.}~\bibnamefont {Yamaura}}, \bibinfo {author}
  {\bibfnamefont {P.~G.}\ \bibnamefont {Radaelli}}, \bibinfo {author}
  {\bibfnamefont {A.~A.}\ \bibnamefont {Belik}},\ and\ \bibinfo {author}
  {\bibfnamefont {R.~D.}\ \bibnamefont {Johnson}},\ }\bibfield  {title}
  {\bibinfo {title} {Magnetic structure and spin-flop transition in the
  $a$-site columnar-ordered quadruple perovskite
  ${\mathrm{tmmn}}_{3}{\mathrm{o}}_{6}$},\ }\href
  {https://doi.org/10.1103/PhysRevB.99.104424} {\bibfield  {journal} {\bibinfo
  {journal} {Phys. Rev. B}\ }\textbf {\bibinfo {volume} {99}},\ \bibinfo
  {pages} {104424} (\bibinfo {year} {2019})}\BibitemShut {NoStop}%
\bibitem [{\citenamefont {Clark}\ and\ \citenamefont
  {Callen}(1968)}]{Clark1968}%
  \BibitemOpen
  \bibfield  {author} {\bibinfo {author} {\bibfnamefont {A.~E.}\ \bibnamefont
  {Clark}}\ and\ \bibinfo {author} {\bibfnamefont {E.}~\bibnamefont {Callen}},\
  }\bibfield  {title} {\bibinfo {title} {Neel ferrimagnets in large magnetic
  fields},\ }\href@noop {} {\bibfield  {journal} {\bibinfo  {journal} {J. Appl.
  Phys.}\ }\textbf {\bibinfo {volume} {39}},\ \bibinfo {pages} {5972} (\bibinfo
  {year} {1968})}\BibitemShut {NoStop}%
\bibitem [{\citenamefont {Ishimatsu}\ \emph {et~al.}(1999)\citenamefont
  {Ishimatsu}, \citenamefont {Hashizume}, \citenamefont {Hamada}, \citenamefont
  {Hosoito}, \citenamefont {Nelson}, \citenamefont {Venkataraman},
  \citenamefont {Srajer},\ and\ \citenamefont {Lang}}]{IshimatsuPRB1999}%
  \BibitemOpen
  \bibfield  {author} {\bibinfo {author} {\bibfnamefont {N.}~\bibnamefont
  {Ishimatsu}}, \bibinfo {author} {\bibfnamefont {H.}~\bibnamefont
  {Hashizume}}, \bibinfo {author} {\bibfnamefont {S.}~\bibnamefont {Hamada}},
  \bibinfo {author} {\bibfnamefont {N.}~\bibnamefont {Hosoito}}, \bibinfo
  {author} {\bibfnamefont {C.~S.}\ \bibnamefont {Nelson}}, \bibinfo {author}
  {\bibfnamefont {C.~T.}\ \bibnamefont {Venkataraman}}, \bibinfo {author}
  {\bibfnamefont {G.}~\bibnamefont {Srajer}},\ and\ \bibinfo {author}
  {\bibfnamefont {J.~C.}\ \bibnamefont {Lang}},\ }\bibfield  {title} {\bibinfo
  {title} {Magnetic structure of fe/gd multilayers determined by resonant x-ray
  magnetic scattering},\ }\href@noop {} {\bibfield  {journal} {\bibinfo
  {journal} {Phys. Rev. B}\ }\textbf {\bibinfo {volume} {60}},\ \bibinfo
  {pages} {9596} (\bibinfo {year} {1999})}\BibitemShut {NoStop}%
\bibitem [{\citenamefont {Haskel}\ \emph {et~al.}(2001)\citenamefont {Haskel},
  \citenamefont {Srajer}, \citenamefont {Lang}, \citenamefont {Pollmann},
  \citenamefont {Nelson}, \citenamefont {Jiang},\ and\ \citenamefont
  {Bader}}]{HaskelPRL2001}%
  \BibitemOpen
  \bibfield  {author} {\bibinfo {author} {\bibfnamefont {D.}~\bibnamefont
  {Haskel}}, \bibinfo {author} {\bibfnamefont {G.}~\bibnamefont {Srajer}},
  \bibinfo {author} {\bibfnamefont {J.~C.}\ \bibnamefont {Lang}}, \bibinfo
  {author} {\bibfnamefont {J.}~\bibnamefont {Pollmann}}, \bibinfo {author}
  {\bibfnamefont {C.~S.}\ \bibnamefont {Nelson}}, \bibinfo {author}
  {\bibfnamefont {J.~S.}\ \bibnamefont {Jiang}},\ and\ \bibinfo {author}
  {\bibfnamefont {S.~D.}\ \bibnamefont {Bader}},\ }\bibfield  {title} {\bibinfo
  {title} {Enhanced interfacial magnetic coupling of gd $/$fe multilayers},\
  }\href@noop {} {\bibfield  {journal} {\bibinfo  {journal} {Phys. Rev. Lett.}\
  }\textbf {\bibinfo {volume} {87}},\ \bibinfo {pages} {207201} (\bibinfo
  {year} {2001})}\BibitemShut {NoStop}%
\bibitem [{\citenamefont {Montoya}\ \emph
  {et~al.}(2017{\natexlab{a}})\citenamefont {Montoya}, \citenamefont {Couture},
  \citenamefont {Chess}, \citenamefont {Lee}, \citenamefont {Kent},
  \citenamefont {Henze}, \citenamefont {Sinha}, \citenamefont {Im},
  \citenamefont {Kevan}, \citenamefont {Fischer}, \citenamefont {McMorran},
  \citenamefont {Lomakin}, \citenamefont {Roy},\ and\ \citenamefont
  {Fullerton}}]{Montoya1PRB17}%
  \BibitemOpen
  \bibfield  {author} {\bibinfo {author} {\bibfnamefont {S.~A.}\ \bibnamefont
  {Montoya}}, \bibinfo {author} {\bibfnamefont {S.}~\bibnamefont {Couture}},
  \bibinfo {author} {\bibfnamefont {J.~J.}\ \bibnamefont {Chess}}, \bibinfo
  {author} {\bibfnamefont {J.~C.~T.}\ \bibnamefont {Lee}}, \bibinfo {author}
  {\bibfnamefont {N.}~\bibnamefont {Kent}}, \bibinfo {author} {\bibfnamefont
  {D.}~\bibnamefont {Henze}}, \bibinfo {author} {\bibfnamefont {S.~K.}\
  \bibnamefont {Sinha}}, \bibinfo {author} {\bibfnamefont {M.-Y.}\ \bibnamefont
  {Im}}, \bibinfo {author} {\bibfnamefont {S.~D.}\ \bibnamefont {Kevan}},
  \bibinfo {author} {\bibfnamefont {P.}~\bibnamefont {Fischer}}, \bibinfo
  {author} {\bibfnamefont {B.~J.}\ \bibnamefont {McMorran}}, \bibinfo {author}
  {\bibfnamefont {V.}~\bibnamefont {Lomakin}}, \bibinfo {author} {\bibfnamefont
  {S.}~\bibnamefont {Roy}},\ and\ \bibinfo {author} {\bibfnamefont {E.~E.}\
  \bibnamefont {Fullerton}},\ }\bibfield  {title} {\bibinfo {title} {Tailoring
  magnetic energies to form dipole skyrmions and skyrmion lattices},\ }\href
  {https://link.aps.org/doi/10.1103/PhysRevB.95.024415} {\bibfield  {journal}
  {\bibinfo  {journal} {Phys. Rev. B}\ }\textbf {\bibinfo {volume} {95}},\
  \bibinfo {pages} {024415} (\bibinfo {year} {2017}{\natexlab{a}})}\BibitemShut
  {NoStop}%
\bibitem [{\citenamefont {Montoya}\ \emph
  {et~al.}(2017{\natexlab{b}})\citenamefont {Montoya}, \citenamefont {Couture},
  \citenamefont {Chess}, \citenamefont {Lee}, \citenamefont {Kent},
  \citenamefont {Im}, \citenamefont {Kevan}, \citenamefont {Fischer},
  \citenamefont {McMorran}, \citenamefont {Roy}, \citenamefont {Lomakin},\ and\
  \citenamefont {Fullerton}}]{Montoya2PRB17}%
  \BibitemOpen
  \bibfield  {author} {\bibinfo {author} {\bibfnamefont {S.~A.}\ \bibnamefont
  {Montoya}}, \bibinfo {author} {\bibfnamefont {S.}~\bibnamefont {Couture}},
  \bibinfo {author} {\bibfnamefont {J.~J.}\ \bibnamefont {Chess}}, \bibinfo
  {author} {\bibfnamefont {J.~C.~T.}\ \bibnamefont {Lee}}, \bibinfo {author}
  {\bibfnamefont {N.}~\bibnamefont {Kent}}, \bibinfo {author} {\bibfnamefont
  {M.-Y.}\ \bibnamefont {Im}}, \bibinfo {author} {\bibfnamefont {S.~D.}\
  \bibnamefont {Kevan}}, \bibinfo {author} {\bibfnamefont {P.}~\bibnamefont
  {Fischer}}, \bibinfo {author} {\bibfnamefont {B.~J.}\ \bibnamefont
  {McMorran}}, \bibinfo {author} {\bibfnamefont {S.}~\bibnamefont {Roy}},
  \bibinfo {author} {\bibfnamefont {V.}~\bibnamefont {Lomakin}},\ and\ \bibinfo
  {author} {\bibfnamefont {E.~E.}\ \bibnamefont {Fullerton}},\ }\bibfield
  {title} {\bibinfo {title} {Resonant properties of dipole skyrmions in
  amorphous fe/gd multilayers},\ }\href
  {https://link.aps.org/doi/10.1103/PhysRevB.95.224405} {\bibfield  {journal}
  {\bibinfo  {journal} {Phys. Rev. B}\ }\textbf {\bibinfo {volume} {95}},\
  \bibinfo {pages} {224405} (\bibinfo {year} {2017}{\natexlab{b}})}\BibitemShut
  {NoStop}%
\bibitem [{\citenamefont {Takanashi}\ \emph {et~al.}(1992)\citenamefont
  {Takanashi}, \citenamefont {Kamiguchi}, \citenamefont {Fujimori},\ and\
  \citenamefont {Motokawa}}]{Takanashi1992}%
  \BibitemOpen
  \bibfield  {author} {\bibinfo {author} {\bibfnamefont {K.}~\bibnamefont
  {Takanashi}}, \bibinfo {author} {\bibfnamefont {Y.}~\bibnamefont
  {Kamiguchi}}, \bibinfo {author} {\bibfnamefont {H.}~\bibnamefont
  {Fujimori}},\ and\ \bibinfo {author} {\bibfnamefont {M.}~\bibnamefont
  {Motokawa}},\ }\bibfield  {title} {\bibinfo {title} {Magnetization and
  magnetoresistance of fe/gd ferrimagnetic multilayer films},\ }\href@noop {}
  {\bibfield  {journal} {\bibinfo  {journal} {J. Phys. Soc. Japan}\ }\textbf
  {\bibinfo {volume} {61}},\ \bibinfo {pages} {3721} (\bibinfo {year}
  {1992})}\BibitemShut {NoStop}%
\bibitem [{\citenamefont {Baczewski}\ \emph {et~al.}(1998)\citenamefont
  {Baczewski}, \citenamefont {R.~Kalinowski},\ and\ \citenamefont
  {Wawro}}]{Baczewski}%
  \BibitemOpen
  \bibfield  {author} {\bibinfo {author} {\bibfnamefont {L.~T.}\ \bibnamefont
  {Baczewski}}, \bibinfo {author} {\bibfnamefont {R.}~\bibnamefont
  {R.~Kalinowski}},\ and\ \bibinfo {author} {\bibfnamefont {A.}~\bibnamefont
  {Wawro}},\ }\bibfield  {title} {\bibinfo {title} {Magnetization and
  anisotropy in fe/gd multilayers},\ }\href@noop {} {\bibfield  {journal}
  {\bibinfo  {journal} {J. Magn. Magn. Mater.}\ }\textbf {\bibinfo {volume}
  {177}},\ \bibinfo {pages} {1305} (\bibinfo {year} {1998})}\BibitemShut
  {NoStop}%
\bibitem [{\citenamefont {Kamiguchi}\ \emph {et~al.}(1989)\citenamefont
  {Kamiguchi}, \citenamefont {Hayakawa},\ and\ \citenamefont
  {Fujimori}}]{Kamiguchi89}%
  \BibitemOpen
  \bibfield  {author} {\bibinfo {author} {\bibfnamefont {Y.}~\bibnamefont
  {Kamiguchi}}, \bibinfo {author} {\bibfnamefont {Y.}~\bibnamefont
  {Hayakawa}},\ and\ \bibinfo {author} {\bibfnamefont {H.}~\bibnamefont
  {Fujimori}},\ }\bibfield  {title} {\bibinfo {title} {Anomalous field
  dependence of magnetoresistance in fe/gd multilayered ferrimagnets},\
  }\href@noop {} {\bibfield  {journal} {\bibinfo  {journal} {Appl. Phys.
  Lett.}\ }\textbf {\bibinfo {volume} {55}},\ \bibinfo {pages} {1918} (\bibinfo
  {year} {1989})}\BibitemShut {NoStop}%
\bibitem [{\citenamefont {Drovosekov}\ \emph {et~al.}(2015)\citenamefont
  {Drovosekov}, \citenamefont {Kreines}, \citenamefont {Savitsky},
  \citenamefont {Kravtsov}, \citenamefont {Blagodatkov}, \citenamefont
  {Ryabukhina}, \citenamefont {Milyaev}, \citenamefont {Ustinov}, \citenamefont
  {Pashaev}, \citenamefont {Subbotin},\ and\ \citenamefont
  {Prutskov}}]{Drovosekov15}%
  \BibitemOpen
  \bibfield  {author} {\bibinfo {author} {\bibfnamefont {A.~B.}\ \bibnamefont
  {Drovosekov}}, \bibinfo {author} {\bibfnamefont {N.~M.}\ \bibnamefont
  {Kreines}}, \bibinfo {author} {\bibfnamefont {A.~O.}\ \bibnamefont
  {Savitsky}}, \bibinfo {author} {\bibfnamefont {E.~A.}\ \bibnamefont
  {Kravtsov}}, \bibinfo {author} {\bibfnamefont {D.~V.}\ \bibnamefont
  {Blagodatkov}}, \bibinfo {author} {\bibfnamefont {M.~V.}\ \bibnamefont
  {Ryabukhina}}, \bibinfo {author} {\bibfnamefont {M.~A.}\ \bibnamefont
  {Milyaev}}, \bibinfo {author} {\bibfnamefont {V.~V.}\ \bibnamefont
  {Ustinov}}, \bibinfo {author} {\bibfnamefont {E.~M.}\ \bibnamefont
  {Pashaev}}, \bibinfo {author} {\bibfnamefont {I.~A.}\ \bibnamefont
  {Subbotin}},\ and\ \bibinfo {author} {\bibfnamefont {G.~V.}\ \bibnamefont
  {Prutskov}},\ }\bibfield  {title} {\bibinfo {title} {Interlayer coupling in
  fe/cr/gd multilayer structures},\ }\href@noop {} {\bibfield  {journal}
  {\bibinfo  {journal} {J. Exp. Theor.}\ }\textbf {\bibinfo {volume} {120}},\
  \bibinfo {pages} {1041} (\bibinfo {year} {2015})}\BibitemShut {NoStop}%
\bibitem [{\citenamefont {Drovosekov}\ \emph {et~al.}(2018)\citenamefont
  {Drovosekov}, \citenamefont {Ryabukhina}, \citenamefont {Kholin},
  \citenamefont {Kreines}, \citenamefont {Manuilovich}, \citenamefont
  {Savitsky}, \citenamefont {Kravtsov}, \citenamefont {Proglyado},
  \citenamefont {Ustinov}, \citenamefont {Keller}, \citenamefont {Khaydukov},
  \citenamefont {Choi},\ and\ \citenamefont {Haskel}}]{Drovosekov18}%
  \BibitemOpen
  \bibfield  {author} {\bibinfo {author} {\bibfnamefont {A.~B.}\ \bibnamefont
  {Drovosekov}}, \bibinfo {author} {\bibfnamefont {M.~V.}\ \bibnamefont
  {Ryabukhina}}, \bibinfo {author} {\bibfnamefont {D.~I.}\ \bibnamefont
  {Kholin}}, \bibinfo {author} {\bibfnamefont {N.~M.}\ \bibnamefont {Kreines}},
  \bibinfo {author} {\bibfnamefont {E.~A.}\ \bibnamefont {Manuilovich}},
  \bibinfo {author} {\bibfnamefont {A.~O.}\ \bibnamefont {Savitsky}}, \bibinfo
  {author} {\bibfnamefont {E.~A.}\ \bibnamefont {Kravtsov}}, \bibinfo {author}
  {\bibfnamefont {V.~V.}\ \bibnamefont {Proglyado}}, \bibinfo {author}
  {\bibfnamefont {V.~V.}\ \bibnamefont {Ustinov}}, \bibinfo {author}
  {\bibfnamefont {T.}~\bibnamefont {Keller}}, \bibinfo {author} {\bibfnamefont
  {Y.~N.}\ \bibnamefont {Khaydukov}}, \bibinfo {author} {\bibfnamefont
  {Y.}~\bibnamefont {Choi}},\ and\ \bibinfo {author} {\bibfnamefont
  {D.}~\bibnamefont {Haskel}},\ }\bibfield  {title} {\bibinfo {title} {Effect
  of cr spacer on structural and magnetic properties of fe/gd multilayers},\
  }\href@noop {} {\bibfield  {journal} {\bibinfo  {journal} {J. Exp. Theor.}\
  }\textbf {\bibinfo {volume} {127}},\ \bibinfo {pages} {742} (\bibinfo {year}
  {2018})}\BibitemShut {NoStop}%
\bibitem [{\citenamefont {Takanashi}\ \emph {et~al.}(1993)\citenamefont
  {Takanashi}, \citenamefont {Kurokawa},\ and\ \citenamefont
  {Fujimori}}]{Takanashi1993}%
  \BibitemOpen
  \bibfield  {author} {\bibinfo {author} {\bibfnamefont {K.}~\bibnamefont
  {Takanashi}}, \bibinfo {author} {\bibfnamefont {H.}~\bibnamefont
  {Kurokawa}},\ and\ \bibinfo {author} {\bibfnamefont {H.}~\bibnamefont
  {Fujimori}},\ }\bibfield  {title} {\bibinfo {title} {A novel hysteresis loop
  and indirect exchange coupling in co/pt/gd/pt multilayer films},\ }\href@noop
  {} {\bibfield  {journal} {\bibinfo  {journal} {Appl. Phys. Lett.}\ }\textbf
  {\bibinfo {volume} {63}},\ \bibinfo {pages} {1585} (\bibinfo {year}
  {1993})}\BibitemShut {NoStop}%
\bibitem [{\citenamefont {Merenkov}\ \emph {et~al.}(2001)\citenamefont
  {Merenkov}, \citenamefont {Chizhik}, \citenamefont {Gnatchenko},
  \citenamefont {Baran}, \citenamefont {Szymczak}, \citenamefont
  {Vas'kovskiy},\ and\ \citenamefont {Svalov}}]{Merenkov2001}%
  \BibitemOpen
  \bibfield  {author} {\bibinfo {author} {\bibfnamefont {D.~N.}\ \bibnamefont
  {Merenkov}}, \bibinfo {author} {\bibfnamefont {A.~B.}\ \bibnamefont
  {Chizhik}}, \bibinfo {author} {\bibfnamefont {S.~L.}\ \bibnamefont
  {Gnatchenko}}, \bibinfo {author} {\bibfnamefont {M.}~\bibnamefont {Baran}},
  \bibinfo {author} {\bibfnamefont {R.}~\bibnamefont {Szymczak}}, \bibinfo
  {author} {\bibfnamefont {V.~O.}\ \bibnamefont {Vas'kovskiy}},\ and\ \bibinfo
  {author} {\bibfnamefont {A.~V.}\ \bibnamefont {Svalov}},\ }\bibfield  {title}
  {\bibinfo {title} {H{\textendash}t phase diagram of a multilayered gd/si/co
  film with ferrimagnetic ordering of the layers},\ }\href@noop {} {\bibfield
  {journal} {\bibinfo  {journal} {Low Temp. Phys.}\ }\textbf {\bibinfo {volume}
  {27}},\ \bibinfo {pages} {137} (\bibinfo {year} {2001})}\BibitemShut
  {NoStop}%
\bibitem [{\citenamefont {te~Velthuis}\ \emph {et~al.}(2002)\citenamefont
  {te~Velthuis}, \citenamefont {Jiang}, \citenamefont {Bader},\ and\
  \citenamefont {Felcher}}]{Velthuis02}%
  \BibitemOpen
  \bibfield  {author} {\bibinfo {author} {\bibfnamefont {S.~G.~E.}\
  \bibnamefont {te~Velthuis}}, \bibinfo {author} {\bibfnamefont {J.~S.}\
  \bibnamefont {Jiang}}, \bibinfo {author} {\bibfnamefont {S.~D.}\ \bibnamefont
  {Bader}},\ and\ \bibinfo {author} {\bibfnamefont {G.~P.}\ \bibnamefont
  {Felcher}},\ }\bibfield  {title} {\bibinfo {title} {Spin flop transition in a
  finite antiferromagnetic superlattice: Evolution of the magnetic structure},\
  }\href@noop {} {\bibfield  {journal} {\bibinfo  {journal} {Phys. Rev. Lett.}\
  }\textbf {\bibinfo {volume} {89}},\ \bibinfo {pages} {127203} (\bibinfo
  {year} {2002})}\BibitemShut {NoStop}%
\bibitem [{\citenamefont {Lauter-Pasyuk}\ \emph {et~al.}(2002)\citenamefont
  {Lauter-Pasyuk}, \citenamefont {Lauter}, \citenamefont {Toperverg},
  \citenamefont {Romashev},\ and\ \citenamefont {Ustinov}}]{Pasyuk02}%
  \BibitemOpen
  \bibfield  {author} {\bibinfo {author} {\bibfnamefont {V.}~\bibnamefont
  {Lauter-Pasyuk}}, \bibinfo {author} {\bibfnamefont {H.~J.}\ \bibnamefont
  {Lauter}}, \bibinfo {author} {\bibfnamefont {B.~P.}\ \bibnamefont
  {Toperverg}}, \bibinfo {author} {\bibfnamefont {L.}~\bibnamefont
  {Romashev}},\ and\ \bibinfo {author} {\bibfnamefont {V.}~\bibnamefont
  {Ustinov}},\ }\bibfield  {title} {\bibinfo {title} {Transverse and lateral
  structure of the spin-flop phase in
  $\mathrm{F}\mathrm{e}/\mathrm{C}\mathrm{r}$ antiferromagnetic
  superlattices},\ }\href@noop {} {\bibfield  {journal} {\bibinfo  {journal}
  {Phys. Rev. Lett.}\ }\textbf {\bibinfo {volume} {89}},\ \bibinfo {pages}
  {167203} (\bibinfo {year} {2002})}\BibitemShut {NoStop}%
\bibitem [{\citenamefont {Nagy}\ \emph {et~al.}(2002)\citenamefont {Nagy},
  \citenamefont {Botty\'an}, \citenamefont {Croonenborghs}, \citenamefont
  {De\'ak}, \citenamefont {Degroote}, \citenamefont {Dekoster}, \citenamefont
  {Lauter}, \citenamefont {Lauter-Pasyuk}, \citenamefont {Leupold},
  \citenamefont {Major}, \citenamefont {Meersschaut}, \citenamefont {Nikonov},
  \citenamefont {Petrenko}, \citenamefont {R\"uffer}, \citenamefont
  {Spiering},\ and\ \citenamefont {Szil\'agyi}}]{Nagy02}%
  \BibitemOpen
  \bibfield  {author} {\bibinfo {author} {\bibfnamefont {D.~L.}\ \bibnamefont
  {Nagy}}, \bibinfo {author} {\bibfnamefont {L.}~\bibnamefont {Botty\'an}},
  \bibinfo {author} {\bibfnamefont {B.}~\bibnamefont {Croonenborghs}}, \bibinfo
  {author} {\bibfnamefont {L.}~\bibnamefont {De\'ak}}, \bibinfo {author}
  {\bibfnamefont {B.}~\bibnamefont {Degroote}}, \bibinfo {author}
  {\bibfnamefont {J.}~\bibnamefont {Dekoster}}, \bibinfo {author}
  {\bibfnamefont {H.~J.}\ \bibnamefont {Lauter}}, \bibinfo {author}
  {\bibfnamefont {V.}~\bibnamefont {Lauter-Pasyuk}}, \bibinfo {author}
  {\bibfnamefont {O.}~\bibnamefont {Leupold}}, \bibinfo {author} {\bibfnamefont
  {M.}~\bibnamefont {Major}}, \bibinfo {author} {\bibfnamefont
  {J.}~\bibnamefont {Meersschaut}}, \bibinfo {author} {\bibfnamefont
  {O.}~\bibnamefont {Nikonov}}, \bibinfo {author} {\bibfnamefont
  {A.}~\bibnamefont {Petrenko}}, \bibinfo {author} {\bibfnamefont
  {R.}~\bibnamefont {R\"uffer}}, \bibinfo {author} {\bibfnamefont
  {H.}~\bibnamefont {Spiering}},\ and\ \bibinfo {author} {\bibfnamefont
  {E.}~\bibnamefont {Szil\'agyi}},\ }\bibfield  {title} {\bibinfo {title}
  {Coarsening of antiferromagnetic domains in multilayers: The key role of
  magnetocrystalline anisotropy},\ }\href
  {https://link.aps.org/doi/10.1103/PhysRevLett.88.157202} {\bibfield
  {journal} {\bibinfo  {journal} {Phys. Rev. Lett.}\ }\textbf {\bibinfo
  {volume} {88}},\ \bibinfo {pages} {157202} (\bibinfo {year}
  {2002})}\BibitemShut {NoStop}%
\bibitem [{\citenamefont {Antropov}\ \emph {et~al.}(2019)\citenamefont
  {Antropov}, \citenamefont {Khaydukov}, \citenamefont {Kravtsov},
  \citenamefont {Makarova},\ and\ \citenamefont {Ustinov}}]{ANTROPOV}%
  \BibitemOpen
  \bibfield  {author} {\bibinfo {author} {\bibfnamefont {N.~O.}\ \bibnamefont
  {Antropov}}, \bibinfo {author} {\bibfnamefont {Y.~N.}\ \bibnamefont
  {Khaydukov}}, \bibinfo {author} {\bibfnamefont {E.~A.}\ \bibnamefont
  {Kravtsov}}, \bibinfo {author} {\bibfnamefont {V.~V.}\ \bibnamefont
  {Makarova}, \bibfnamefont {M.~V.~Progliado}},\ and\ \bibinfo {author}
  {\bibfnamefont {V.~V.}\ \bibnamefont {Ustinov}},\ }\bibfield  {title}
  {\bibinfo {title} {Transition in a magnetic non-collinear spin-flop state in
  a fe/pd/gd/pd superlattice},\ }\href@noop {} {\bibfield  {journal} {\bibinfo
  {journal} {JETP Lett.}\ }\textbf {\bibinfo {volume} {109}},\ \bibinfo {pages}
  {406} (\bibinfo {year} {2019})}\BibitemShut {NoStop}%
\bibitem [{\citenamefont {Pashaev}\ \emph {et~al.}(2020)\citenamefont
  {Pashaev}, \citenamefont {Vasiliev}, \citenamefont {Subbotin}, \citenamefont
  {Prutskov}, \citenamefont {Chesnokov}, \citenamefont {Kovalchuk},
  \citenamefont {Antropov}, \citenamefont {Kravtsov}, \citenamefont
  {Proglyado},\ and\ \citenamefont {Ustinov}}]{Pashaev20}%
  \BibitemOpen
  \bibfield  {author} {\bibinfo {author} {\bibfnamefont {E.}~\bibnamefont
  {Pashaev}}, \bibinfo {author} {\bibfnamefont {A.}~\bibnamefont {Vasiliev}},
  \bibinfo {author} {\bibfnamefont {I.}~\bibnamefont {Subbotin}}, \bibinfo
  {author} {\bibfnamefont {G.}~\bibnamefont {Prutskov}}, \bibinfo {author}
  {\bibfnamefont {Y.~M.}\ \bibnamefont {Chesnokov}}, \bibinfo {author}
  {\bibfnamefont {M.}~\bibnamefont {Kovalchuk}}, \bibinfo {author}
  {\bibfnamefont {N.}~\bibnamefont {Antropov}}, \bibinfo {author}
  {\bibfnamefont {E.}~\bibnamefont {Kravtsov}}, \bibinfo {author}
  {\bibfnamefont {V.}~\bibnamefont {Proglyado}},\ and\ \bibinfo {author}
  {\bibfnamefont {V.}~\bibnamefont {Ustinov}},\ }\bibfield  {title} {\bibinfo
  {title} {Analysis of structural features of periodic fe/pd/gd/pd multilayered
  systems},\ }\href@noop {} {\bibfield  {journal} {\bibinfo  {journal}
  {Crystallography Reports}\ }\textbf {\bibinfo {volume} {65}},\ \bibinfo
  {pages} {985} (\bibinfo {year} {2020})}\BibitemShut {NoStop}%
\bibitem [{\citenamefont {Solignac}\ \emph {et~al.}(2012)\citenamefont
  {Solignac}, \citenamefont {Guerrero}, \citenamefont {Gogol}, \citenamefont
  {Maroutian}, \citenamefont {Ott}, \citenamefont {Largeau}, \citenamefont
  {Lecoeur},\ and\ \citenamefont {Pannetier-Lecoeur}}]{Solignac12}%
  \BibitemOpen
  \bibfield  {author} {\bibinfo {author} {\bibfnamefont {A.}~\bibnamefont
  {Solignac}}, \bibinfo {author} {\bibfnamefont {R.}~\bibnamefont {Guerrero}},
  \bibinfo {author} {\bibfnamefont {P.}~\bibnamefont {Gogol}}, \bibinfo
  {author} {\bibfnamefont {T.}~\bibnamefont {Maroutian}}, \bibinfo {author}
  {\bibfnamefont {F.}~\bibnamefont {Ott}}, \bibinfo {author} {\bibfnamefont
  {L.}~\bibnamefont {Largeau}}, \bibinfo {author} {\bibfnamefont
  {P.}~\bibnamefont {Lecoeur}},\ and\ \bibinfo {author} {\bibfnamefont
  {M.}~\bibnamefont {Pannetier-Lecoeur}},\ }\bibfield  {title} {\bibinfo
  {title} {Dual antiferromagnetic coupling at
  ${\mathrm{la}}_{0.67}{\mathrm{sr}}_{0.33}{\mathrm{mno}}_{3}/{\mathrm{srruo}}_{3}$
  interfaces},\ }\href@noop {} {\bibfield  {journal} {\bibinfo  {journal}
  {Phys. Rev. Lett.}\ }\textbf {\bibinfo {volume} {109}},\ \bibinfo {pages}
  {027201} (\bibinfo {year} {2012})}\BibitemShut {NoStop}%
\bibitem [{\citenamefont {Suciu}\ \emph {et~al.}(2002)\citenamefont {Suciu},
  \citenamefont {Toussaint},\ and\ \citenamefont {Voiron}}]{Suciu02}%
  \BibitemOpen
  \bibfield  {author} {\bibinfo {author} {\bibfnamefont {G.}~\bibnamefont
  {Suciu}}, \bibinfo {author} {\bibfnamefont {J.}~\bibnamefont {Toussaint}},\
  and\ \bibinfo {author} {\bibfnamefont {J.}~\bibnamefont {Voiron}},\
  }\bibfield  {title} {\bibinfo {title} {4f{\textendash}3d exchange coupling in
  gd/x/co (x=pt, cr) multilayers},\ }\href@noop {} {\bibfield  {journal}
  {\bibinfo  {journal} {J. Magn. Magn. Mater.}\ }\textbf {\bibinfo {volume}
  {240}},\ \bibinfo {pages} {229} (\bibinfo {year} {2002})}\BibitemShut
  {NoStop}%
\bibitem [{\citenamefont {Khaydukov}\ \emph {et~al.}(2018)\citenamefont
  {Khaydukov}, \citenamefont {Vasenko}, \citenamefont {Kravtsov}, \citenamefont
  {Progliado}, \citenamefont {Zhaketov}, \citenamefont {Csik}, \citenamefont
  {Nikitenko}, \citenamefont {Petrenko}, \citenamefont {Keller}, \citenamefont
  {Golubov}, \citenamefont {Kupriyanov}, \citenamefont {Ustinov}, \citenamefont
  {Aksenov},\ and\ \citenamefont {Keimer}}]{KhaydukovPRB18}%
  \BibitemOpen
  \bibfield  {author} {\bibinfo {author} {\bibfnamefont {Y.~N.}\ \bibnamefont
  {Khaydukov}}, \bibinfo {author} {\bibfnamefont {A.~S.}\ \bibnamefont
  {Vasenko}}, \bibinfo {author} {\bibfnamefont {E.~A.}\ \bibnamefont
  {Kravtsov}}, \bibinfo {author} {\bibfnamefont {V.~V.}\ \bibnamefont
  {Progliado}}, \bibinfo {author} {\bibfnamefont {V.~D.}\ \bibnamefont
  {Zhaketov}}, \bibinfo {author} {\bibfnamefont {A.}~\bibnamefont {Csik}},
  \bibinfo {author} {\bibfnamefont {Y.~V.}\ \bibnamefont {Nikitenko}}, \bibinfo
  {author} {\bibfnamefont {A.~V.}\ \bibnamefont {Petrenko}}, \bibinfo {author}
  {\bibfnamefont {T.}~\bibnamefont {Keller}}, \bibinfo {author} {\bibfnamefont
  {A.~A.}\ \bibnamefont {Golubov}}, \bibinfo {author} {\bibfnamefont {M.~Y.}\
  \bibnamefont {Kupriyanov}}, \bibinfo {author} {\bibfnamefont {V.~V.}\
  \bibnamefont {Ustinov}}, \bibinfo {author} {\bibfnamefont {V.~L.}\
  \bibnamefont {Aksenov}},\ and\ \bibinfo {author} {\bibfnamefont
  {B.}~\bibnamefont {Keimer}},\ }\bibfield  {title} {\bibinfo {title} {Magnetic
  and superconducting phase diagram of nb/gd/nb trilayers},\ }\href
  {https://link.aps.org/doi/10.1103/PhysRevB.97.144511} {\bibfield  {journal}
  {\bibinfo  {journal} {Phys. Rev. B}\ }\textbf {\bibinfo {volume} {97}},\
  \bibinfo {pages} {144511} (\bibinfo {year} {2018})}\BibitemShut {NoStop}%
\bibitem [{\citenamefont {Khaydukov}\ \emph {et~al.}(2019)\citenamefont
  {Khaydukov}, \citenamefont {Kravtsov}, \citenamefont {Zhaketov},
  \citenamefont {Progliado}, \citenamefont {Kim}, \citenamefont {Nikitenko},
  \citenamefont {Keller}, \citenamefont {Ustinov}, \citenamefont {Aksenov},\
  and\ \citenamefont {Keimer}}]{Khaydukov19}%
  \BibitemOpen
  \bibfield  {author} {\bibinfo {author} {\bibfnamefont {Y.~N.}\ \bibnamefont
  {Khaydukov}}, \bibinfo {author} {\bibfnamefont {E.~A.}\ \bibnamefont
  {Kravtsov}}, \bibinfo {author} {\bibfnamefont {V.~D.}\ \bibnamefont
  {Zhaketov}}, \bibinfo {author} {\bibfnamefont {V.~V.}\ \bibnamefont
  {Progliado}}, \bibinfo {author} {\bibfnamefont {G.}~\bibnamefont {Kim}},
  \bibinfo {author} {\bibfnamefont {Y.~V.}\ \bibnamefont {Nikitenko}}, \bibinfo
  {author} {\bibfnamefont {T.}~\bibnamefont {Keller}}, \bibinfo {author}
  {\bibfnamefont {V.~V.}\ \bibnamefont {Ustinov}}, \bibinfo {author}
  {\bibfnamefont {V.~L.}\ \bibnamefont {Aksenov}},\ and\ \bibinfo {author}
  {\bibfnamefont {B.}~\bibnamefont {Keimer}},\ }\bibfield  {title} {\bibinfo
  {title} {Magnetic proximity effect in nb/gd superlattices seen by neutron
  reflectometry},\ }\href {https://doi.org/10.1103/PhysRevB.99.140503}
  {\bibfield  {journal} {\bibinfo  {journal} {Phys. Rev. B}\ }\textbf {\bibinfo
  {volume} {99}},\ \bibinfo {pages} {140503} (\bibinfo {year}
  {2019})}\BibitemShut {NoStop}%
\bibitem [{\citenamefont {Hj\"orvarsson}\ \emph {et~al.}(1997)\citenamefont
  {Hj\"orvarsson}, \citenamefont {Dura}, \citenamefont {Isberg}, \citenamefont
  {Watanabe}, \citenamefont {Udovic}, \citenamefont {Andersson},\ and\
  \citenamefont {Majkrzak}}]{Hjoravrson97}%
  \BibitemOpen
  \bibfield  {author} {\bibinfo {author} {\bibfnamefont {B.}~\bibnamefont
  {Hj\"orvarsson}}, \bibinfo {author} {\bibfnamefont {J.~A.}\ \bibnamefont
  {Dura}}, \bibinfo {author} {\bibfnamefont {P.}~\bibnamefont {Isberg}},
  \bibinfo {author} {\bibfnamefont {T.}~\bibnamefont {Watanabe}}, \bibinfo
  {author} {\bibfnamefont {T.~J.}\ \bibnamefont {Udovic}}, \bibinfo {author}
  {\bibfnamefont {G.}~\bibnamefont {Andersson}},\ and\ \bibinfo {author}
  {\bibfnamefont {C.~F.}\ \bibnamefont {Majkrzak}},\ }\bibfield  {title}
  {\bibinfo {title} {Reversible tuning of the magnetic exchange coupling in
  fe/v (001) superlattices using hydrogen},\ }\href
  {https://link.aps.org/doi/10.1103/PhysRevLett.79.901} {\bibfield  {journal}
  {\bibinfo  {journal} {Phys. Rev. Lett.}\ }\textbf {\bibinfo {volume} {79}},\
  \bibinfo {pages} {901} (\bibinfo {year} {1997})}\BibitemShut {NoStop}%
\bibitem [{\citenamefont {Leiner}\ \emph {et~al.}(2002)\citenamefont {Leiner},
  \citenamefont {Westerholt}, \citenamefont {Hj\"orvarsson},\ and\
  \citenamefont {Zabel}}]{Leiner02}%
  \BibitemOpen
  \bibfield  {author} {\bibinfo {author} {\bibfnamefont {V.}~\bibnamefont
  {Leiner}}, \bibinfo {author} {\bibfnamefont {K.}~\bibnamefont {Westerholt}},
  \bibinfo {author} {\bibfnamefont {B.}~\bibnamefont {Hj\"orvarsson}},\ and\
  \bibinfo {author} {\bibfnamefont {H.}~\bibnamefont {Zabel}},\ }\bibfield
  {title} {\bibinfo {title} {Tunability of the interlayer exchange coupling},\
  }\href {https://doi.org/10.1088%2F0022-3727%2F35%2F19%2F308} {\bibfield
  {journal} {\bibinfo  {journal} {J. Phys. D: Appl. Phys}\ }\textbf {\bibinfo
  {volume} {35}},\ \bibinfo {pages} {2377} (\bibinfo {year}
  {2002})}\BibitemShut {NoStop}%
\bibitem [{\citenamefont {Schreyer}\ \emph {et~al.}(1993)\citenamefont
  {Schreyer}, \citenamefont {Br\"ohl}, \citenamefont {Ankner}, \citenamefont
  {Majkrzak}, \citenamefont {Zeidler}, \citenamefont {B\"odeker}, \citenamefont
  {Metoki},\ and\ \citenamefont {Zabel}}]{Schreyer93}%
  \BibitemOpen
  \bibfield  {author} {\bibinfo {author} {\bibfnamefont {A.}~\bibnamefont
  {Schreyer}}, \bibinfo {author} {\bibfnamefont {K.}~\bibnamefont {Br\"ohl}},
  \bibinfo {author} {\bibfnamefont {J.~F.}\ \bibnamefont {Ankner}}, \bibinfo
  {author} {\bibfnamefont {C.~F.}\ \bibnamefont {Majkrzak}}, \bibinfo {author}
  {\bibfnamefont {T.}~\bibnamefont {Zeidler}}, \bibinfo {author} {\bibfnamefont
  {P.}~\bibnamefont {B\"odeker}}, \bibinfo {author} {\bibfnamefont
  {N.}~\bibnamefont {Metoki}},\ and\ \bibinfo {author} {\bibfnamefont
  {H.}~\bibnamefont {Zabel}},\ }\bibfield  {title} {\bibinfo {title}
  {Oscillatory exchange coupling in co/cu(111) superlattices},\ }\href
  {https://link.aps.org/doi/10.1103/PhysRevB.47.15334} {\bibfield  {journal}
  {\bibinfo  {journal} {Phys. Rev. B}\ }\textbf {\bibinfo {volume} {47}},\
  \bibinfo {pages} {15334} (\bibinfo {year} {1993})}\BibitemShut {NoStop}%
\bibitem [{\citenamefont {Hecker}\ \emph {et~al.}(2005)\citenamefont {Hecker},
  \citenamefont {Valencia}, \citenamefont {Oppeneer}, \citenamefont {Mertins},\
  and\ \citenamefont {Schneider}}]{Hecker05}%
  \BibitemOpen
  \bibfield  {author} {\bibinfo {author} {\bibfnamefont {M.}~\bibnamefont
  {Hecker}}, \bibinfo {author} {\bibfnamefont {S.}~\bibnamefont {Valencia}},
  \bibinfo {author} {\bibfnamefont {P.~M.}\ \bibnamefont {Oppeneer}}, \bibinfo
  {author} {\bibfnamefont {H.-C.}\ \bibnamefont {Mertins}},\ and\ \bibinfo
  {author} {\bibfnamefont {C.~M.}\ \bibnamefont {Schneider}},\ }\bibfield
  {title} {\bibinfo {title} {Polarized soft-x-ray reflection spectroscopy of
  giant magnetoresistive co/cu multilayers},\ }\href@noop {} {\bibfield
  {journal} {\bibinfo  {journal} {Phys. Rev. B}\ }\textbf {\bibinfo {volume}
  {72}},\ \bibinfo {pages} {054437} (\bibinfo {year} {2005})}\BibitemShut
  {NoStop}%
\end{thebibliography}%

\end{document}